\documentclass[10pt,conference]{IEEEtran}
\IEEEoverridecommandlockouts

\usepackage{amsmath,amsfonts}
\usepackage{algorithmic}
\usepackage{graphicx}
\usepackage{textcomp}
\usepackage{xcolor}
\usepackage{booktabs}
\usepackage{multirow}
\usepackage{tabularx}
\usepackage{array}
\usepackage{comment}
\usepackage{balance}
\usepackage[nobreak,space,compress]{cite}
\usepackage{hyperref}            
\usepackage{xurl}                
\hypersetup{hidelinks,breaklinks=true}
\usepackage{tikz}
\usetikzlibrary{arrows.meta,positioning}
\usepackage{hyperref}
\usepackage[T1]{fontenc}
\usepackage[utf8]{inputenc}
\usepackage{caption}
\usepackage{enumitem}
\usepackage{makecell} 

\def\BibTeX{{\rm B\kern-.05em{\sc i\kern-.025em b}\kern-.08em
    T\kern-.1667em\lower.7ex\hbox{E}\kern-.125emX}}

\let\oldref\ref
\renewcommand{\ref}[1]{
  \tikz[baseline=(X.base)] \node[draw=red, rounded corners=2pt, inner sep=1pt](X){\oldref{#1}};
}

\newcolumntype{A}{>{\centering\arraybackslash}p{0.9em}}

\usepackage{mdframed}
\newmdenv[
  linewidth=0.4pt,
  roundcorner=2pt,
  backgroundcolor=gray!5,
  skipabove=0pt,
  skipbelow=0pt,
  innertopmargin=4pt, innerbottommargin=4pt,
  innerleftmargin=6pt, innerrightmargin=6pt
]{rqbox}

\begin{document}

\title{Android Instrumentation Testing in Continuous Integration: Practices, Patterns, and Performance}

\author
{
  \IEEEauthorblockN{Hamid Parsazadeh}
  \IEEEauthorblockA{Faculty of Information\\
  University of Toronto\\
  Toronto, Ontario, Canada\\
  b.parsazadeh@mail.utoronto.ca
  }
  \and
  \IEEEauthorblockN{Taher A. Ghaleb}
  \IEEEauthorblockA{Department of Computer Science\\
  Trent University\\
  Peterborough, Ontario, Canada\\
  taherghaleb@trentu.ca
  }
  \and
  \IEEEauthorblockN{Safwat Hassan}
  \IEEEauthorblockA{Faculty of Information\\
  University of Toronto\\
  Toronto, Ontario, Canada\\
  safwat.hassan@utoronto.ca
  }
}

\maketitle

\begin{abstract}
Android instrumentation tests—end-to-end tests that run on a device or emulator—can catch problems that simpler tests miss. However, running these tests automatically in continuous integration (CI) is often difficult because emulator setup is fragile and configurations tend to drift over time. We study how open-source Android apps run instrumentation tests in CI by analyzing 4,518 repositories that use CI (snapshot: Aug.\ 10, 2025). We examine CI workflow files, scripts, and build configurations to identify cases where device setup is defined in Gradle (e.g., Gradle Managed Devices). Our results answer three questions about adoption, evolution, and outcomes. First, only about one in ten repositories (481/4,518; 10.6\%) run instrumentation tests in CI, typically using either reusable community components or repository-specific custom scripts to set up emulators. Second, these setups usually stay the same over time; when changes happen, projects tend to move from custom scripts toward reusable community components. Third, we study why projects change their CI setup by analyzing their commits, pull requests, and issue messages. We evaluate how different setup styles perform using GitHub Actions run- and step-level metadata (e.g., outcomes, duration, reruns, and queue delay). We find that teams often change approaches to expand test coverage, and that each approach fits different needs: community-based setups are typically the most reliable and efficient for everyday checks on new code, third-party device labs suit scheduled regression testing but can be costlier and fail more often, and custom scripting provides flexibility but is associated with more reruns.
\end{abstract}


\maketitle

\section{Introduction}
\label{sec:introduction}
Instrumentation tests provide end-to-end assurance for Android apps by validating behavior on devices or emulators. Despite their importance, prior research shows that Android projects often struggle to run and maintain such tests effectively at scale, with practical challenges around cost, brittleness, and environmental complexity~\cite{kong2019androidslr,pecorelli2022androidtesting,mahmud2025androidfallsShort}. These challenges become more pronounced in headless continuous integration (CI), where unstable execution environments and configuration drift can slow runs and contribute to nondeterministic outcomes (i.e., flakiness)~\cite{luo2014flaky,Lam2020Lifecycle}.

Although prior studies investigated CI configuration practices and pipeline evolution in open-source software (including Android)~\cite{hilton2016ci,vasilescu2015ci,ghaleb2025cicd,zampetti2021cicd,mazrae2023cicdtools}, there remains a lack of a clear, evidence-based understanding of how Android instrumentation tests are executed in CI, especially concerning which execution environments (emulator/device setups) are used and how these choices evolve over time. CI configuration mining characterizes pipelines at scale but often cannot verify instrumentation-test execution and its underlying environment from repository-auditable evidence, while Android testing research documents testing approaches, quality, adoption, and challenges without directly characterizing observable CI execution environments or how those environment choices evolve over time~\cite{kong2019androidslr,pecorelli2022androidtesting,mahmud2025androidfallsShort}. We bridge this gap by linking CI automation (workflows and invoked artifacts) to build-configuration signals to identify where instrumentation tests run, how execution-environment styles evolve, and what operational trade-offs these styles entail.

To address this, we investigate execution-environment patterns for instrumentation testing through a comprehensive empirical study of 4,518 open-source Android repositories that support continuous integration. The study is structured around three complementary dimensions: \emph{what is used today}, \emph{how it changes over time}, and \emph{why it changes and what these changes imply operationally}. We first establish a baseline by identifying adoption trends, execution-environment styles, and invocation processes. Building on this baseline, we examine the evolution of execution-environment choices on the default branch over time. Finally, we analyze trade-offs among execution styles by relating observed changes to developer-facing motivations and CI operational characteristics.
Specifically, we address the following three research questions (RQs).

\smallskip
\noindent\textbf{RQ1: What are the current adoption practices for instrumentation testing in CI?}
We detect CI execution of instrumentation tests, classify execution-environment styles, and record invocation mechanisms when observable. We observe that instrumentation testing in CI is uncommon (481/4{,}518; 10.6\%), with adoption mainly dominated by \textit{Community} wrappers and \textit{Custom} scripting.

\smallskip
\noindent\textbf{RQ2: How do execution environments evolve over time on the default branch?}
We segment default-branch histories into stable ``episodes'' of unchanged environment styles. We observe that execution-environment configurations change infrequently, and when migrations occur, they most often go from Custom to Community.

\smallskip
\noindent\textbf{RQ3: What drives projects to change instrumentation-test environments, and how does CI performance vary by style?}
We mine episode-boundary development context to infer motivations for environment changes and analyze CI run/step telemetry to compare operational profiles across styles. We observe that teams often change environments to expand or strengthen CI-backed instrumentation testing, and that styles exhibit distinct operational profiles.

\vspace{3pt}
\noindent\textbf{Contributions.} This paper makes the following contributions.
\begin{itemize}
  \item A large-scale, cross-layer analysis of 4,518 CI-enabled Android repositories that detects CI-based instrumentation-test signals from workflows/commands, and mines Gradle build scripts to identify GMD when device setup is declared in Gradle.
    
  \item An execution-environment taxonomy (\emph{where} tests run) covering Community wrappers, Custom scripting, GMD, Third Party device labs, and rare Real Device cases, along with a secondary characterization of invocation mechanisms (\emph{how} tests are triggered) when explicitly observable.

  \item A longitudinal episode-based framework for reconstructing default-branch execution-environment evolution, including state-machine summaries of transitions across styles.

  \item An analysis framework that combines episode-boundary development context (commit/PR/issue text) with CI run- and step-level telemetry to study environment-change drivers and style-specific operational profiles.
      
  \item A replication package (data, scripts, and reproduction instructions) is publicly available.\footnote{\url{https://doi.org/10.6084/m9.figshare.30934721}}
\end{itemize}

\vspace{3pt}
\noindent\textbf{Paper organization.}
Section~\ref{sec:Literature_Review} provides background and reviews the literature.
Section~\ref{sec:methodology} details our methodology.
Section~\ref{sec:results} presents our empirical findings.
Section~\ref{sec:implications} discusses the implications of our findings.
Section~\ref{sec:threats_to_validty} discusses threats.
Section~\ref{sec:conclusion} concludes the paper and outlines future work.

\section{Background and Related Work}
\label{sec:Literature_Review}
This section reviews the background and related work needed to contextualize our study, covering instrumentation testing in Continuous Integration (CI) workflows, execution environments, CI practices in mobile development, and test reliability issues such as flakiness.

\subsection{Android instrumentation testing and CI}
Instrumentation tests validate Android apps end-to-end by exercising the application on a device or emulator using the Android testing framework~\cite{android-instrumented-tests,android-androidjunitrunner}. Compared to unit tests, they depend on the Android runtime and device state, making them more expensive to run and more sensitive to environmental differences. In CI, these tests are executed as part of an automated build-and-test pipeline~\cite{android-test-cli,android-instrumented-tests}. Prior studies show that, despite their practical importance, Android projects often face challenges running end-to-end tests reliably and consistently, due to factors such as execution cost, instability, and environment complexity~\cite{kong2019androidslr,pecorelli2022androidtesting,mahmud2025androidfallsShort}.

\subsection{Execution environments and configuration fidelity}
Execution reliability depends on execution-environment fidelity, such as keeping device/API configurations consistent and minimizing drift across CI runs~\cite{android_emulator_docs}. In practice, projects differ in where the Android runtime runs and how much of the device setup is standardized versus project-specific, which can affect reproducibility, maintenance effort, and failure modes. Android instrumentation tests are executed under several recurring environment approaches, including reusable community automation (e.g., emulator-related actions)~\cite{reactivecircus_action,action_android}, repository-specific setup via custom scripting~\cite{android_emulator_docs}, Gradle Managed Devices (GMD) configured in the build system~\cite{gradle_gmd_docs}, and third-party device labs/device clouds (e.g., Firebase Test Lab)~\cite{google_github_actions_firebase_test_lab,fazzini2020deviceclouds,lin2023devicefarms}; Real-device execution is also possible but less common~\cite{mahmud2025androidfallsShort,fazzini2020deviceclouds,google_github_actions_firebase_test_lab}. 
These options differ in who specifies device/runtime configuration (CI workflow vs.\ build scripts vs.\ external providers)~\cite{github_workflow_syntax,github_actions_building_blocks,reactivecircus_action,action_android,gradle_gmd_docs,google_github_actions_firebase_test_lab,fazzini2020deviceclouds,lin2023devicefarms}, how standardized the setup is~\cite{github_actions_building_blocks,github_composite_actions,github_reuse_workflows,gradle_gmd_docs}, and how observable the configuration is from repository artifacts~\cite{github_actions_building_blocks,github_composite_actions,github_reuse_workflows,github_self_hosted_runners}---dimensions we operationalize as execution-environment styles in Section~\ref{sec:rq1-method}.

\subsection{CI practices in mobile development and workflow indirection}
In GitHub Actions~\cite{github_actions_ga}, automation is often assembled from reusable components such as actions, composite actions, and reusable workflows~\cite{github_workflow_syntax,github_composite_actions,github_reuse_workflows}. As a result, a workflow file may expose only high-level steps while important execution details reside in referenced components. Execution may also depend on self-hosted runners whose environments are customized by the project~\cite{github_self_hosted_runners}. Prior studies of GitHub Actions failures and resource usage suggest that workflow design has operational consequences~\cite{zheng2025actionsfail,bouzenia2024actionsresources}. Because critical setup logic can be hidden behind reusable actions/workflows and repository scripts, analyzing YAML alone can miss where tests actually run. This motivates our CI-layer mining that expands workflows to include inline commands, referenced scripts, and local actions.

\subsection{Test flakiness and configuration drift}
Flaky tests are tests that can pass or fail nondeterministically under the same code and expected behavior, making them costly to diagnose and maintain~\cite{luo2014flaky,Lam2020Lifecycle,fatima2022flakify}. Android and UI-centric testing is particularly exposed to timing, asynchronous behavior, and device-state variability, contributing to developer-facing maintenance burdens~\cite{thorve2018flakyandroid,romano2021uiflaky,pontillo2024mobileflaky}. Recent evidence also shows persistent practical barriers to effective Android testing in open source~\cite{mahmud2025androidfallsShort}. Environment drift and fragile setup can amplify these issues in headless CI~\cite{android_emulator_docs,ghaleb2025cicd}.

\subsection{Comparison with most relevant work}
\label{sec:closest-work}

\noindent\textbf{Studies on CI practices, configuration mining, and CI outcomes.}
Prior studies have investigated CI adoption and its effects in open-source software broadly~\cite{hilton2016ci,vasilescu2015ci,ghaleb2022studying}, including recent work on CI adoption outcomes in Android apps~\cite{zhou2026ciadoptionmobile}, the heterogeneity of CI configurations and workflow complexity~\cite{abrokwah2025empirical,ghaleb2025cicd}, pipeline evolution, generation, and migration~\cite{zampetti2021cicd,mazrae2023cicdtools,ghaleb2025llm4ci,hossain2025cigrate,chopra2025first}, and CI log analysis support~\cite{barnes2026logsieve,barnesTaskAware2025}. Collectively, these studies improve understanding of how CI is adopted, configured, evolved, and analyzed, but they do not directly characterize \emph{where Android instrumentation tests run} in CI using repository-auditable evidence that expands beyond workflow YAML to invoked scripts/actions and build-system device configuration. Our work complements this literature by detecting CI execution of instrumentation tests through workflow-expanded artifacts, introducing a taxonomy of execution-environment styles grounded in observable repository evidence, and modeling how these styles evolve over time.

\vspace{3pt}
\noindent\textbf{Studies on Android testing and instrumentation-related practice.}
Android testing research has examined the area through systematic reviews, large-scale empirical studies, and practitioner-oriented analyses~\cite{kong2019androidslr,pecorelli2022androidtesting,lin2020testautomationandroid,mahmud2024androidtestingpractices,mahmud2025androidfallsShort}. Prior work has also studied challenges closely related to instrumentation-style and UI-driven testing, such as flaky tests~\cite{thorve2018flakyandroid,romano2021uiflaky,pontillo2024mobileflaky} and testing on device clouds or virtual device farms~\cite{fazzini2020deviceclouds,lin2023devicefarms}. In addition, some work has investigated Android regression testing specifically in CI settings~\cite{wang2023androidrtsci}. However, this literature still does not provide a repository-auditable, cross-layer characterization of \emph{observable CI execution environments} for Android instrumentation tests, nor does it explain how those environment choices change longitudinally across repositories. Our study addresses this gap by focusing specifically on CI-executed instrumentation testing, characterizing execution-environment choices, tracing their evolution, and relating them to operational profiles observed in CI.

\section{Methodology}
\label{sec:methodology}
\begin{figure*}[ht]
  \centering
  \includegraphics[width=1.0\linewidth]{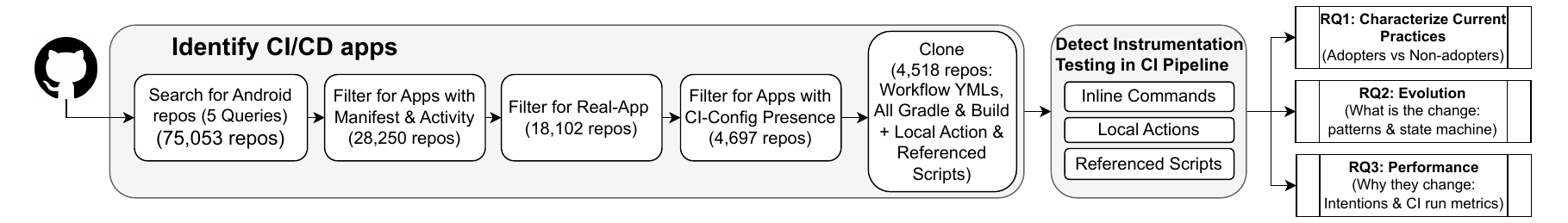}
  \vspace{-6pt}
  \caption{Overview of the dataset construction and analysis pipeline.}
  \label{fig:pipeline}
  \vspace{-5pt}
\end{figure*}

We apply the following filtration and analysis steps to identify repositories that represent Android apps adopting CI services and instrumentation testing in CI. Figure~\ref{fig:pipeline} provides an overview of our approach. The next sections describe our process for identifying Android repositories that adopt instrumentation testing in CI.

\vspace{3pt}
\noindent\textbf{Data collection cut-off.}
All repository metadata and artifacts analyzed in this study are collected as a snapshot on August 10, 2025. Therefore, our results reflect the public state of these repositories and their CI configurations as of that date (and, where applicable, their historical changes up to that cut-off).

\subsection{Identifying CI-adopting apps}
\label{sec:repo-selection}

As shown in Figure~\ref{fig:pipeline}, we progressively filter from a broad set of GitHub candidates to CI-enabled Android app repositories and then clone the final set for analysis.

\vspace{2pt}
\noindent\textbf{Step 1: Searching for candidate repositories.}
We use the GitHub Search API~\cite{github_rest_search_api} to identify candidate repositories via five independent queries: (i) Kotlin projects, (ii) Java projects, (iii) Dart projects, (iv) repositories tagged with the \texttt{android} topic, and (v) repositories in which ``android'' appears in the name, description, or README. Each query applies three constraints: \texttt{stars:>50}, \texttt{fork:false}, and \texttt{archived:false}, which together define the initial target set, yielding 75{,}053 unique repositories.

\vspace{2pt}
\noindent\textbf{Step 2: Filtering repositories with Manifest \& Activity.}
Following prior approaches, we retain repositories that contain an \texttt{AndroidManifest.xml} with at least one \texttt{<activity>} declaration, thereby excluding non-app code (e.g., samples and templates). This step yields 28{,}250 repositories.

\vspace{2pt}
\noindent\textbf{Step 3: Filtering for Real Apps (Removal Keywords).}
Following prior research, we conservatively remove likely non--real-app repositories (e.g., tutorials, samples, templates, SDKs) using keyword-based screening of the repository name, description, README, and selected paths. Flagged repositories are excluded, leaving 18{,}102 repositories.

\vspace{2pt}
\noindent\textbf{Step 4: Filtering for CI Config presence.}
Following prior work on CI configurations, we retain repositories that exhibit CI usage by containing associated configuration files~\cite{ghaleb2025cicd,chopra2025first}. Our detection relies on provider-specific paths and patterns (Table~\ref{tab:ci-services}); repositories without such files are excluded. This step yields 4{,}697 repositories.

\begin{table}[ht]
\centering
\caption{CI services and their Path (regular expression)}
\vspace{-4pt}
\label{tab:ci-services}
\resizebox{\linewidth}{!}{
\begin{tabular}{ll}
\hline
    \textbf{CI service} & \textbf{Path (regular expression)} \\
    \hline
    GitHub Actions  & \texttt{\detokenize{r'\.github/workflows/.*\.(yml|yaml)\$'}} \\
    Travis CI       & \texttt{\detokenize{r'\.travis\.yml\$'}} \\
    CircleCI        & \texttt{\detokenize{r'(\.circleci/config\.yml|circle\.yml)\$'}} \\
    GitLab CI       & \texttt{\detokenize{r'\.gitlab-ci\.yml\$'}} \\
    Azure Pipelines & \texttt{\detokenize{r'azure-pipelines\.yml\$'}} \\
    AppVeyor        & \texttt{\detokenize{r'(\.appveyor\.yml|appveyor\.yml)\$'}} \\
    Bitbucket       & \texttt{\detokenize{r'bitbucket-pipelines\.yml\$'}} \\
    Bitrise         & \texttt{\detokenize{r'bitrise\.yml\$'}} \\
    Bamboo          & \texttt{\detokenize{r'bamboo\.yml\$'}} \\
    Codeship        & \texttt{\detokenize{r'codeship-services\.yml\$'}} \\
    GoCD            & \texttt{\detokenize{r'\.gocd\.yaml\$'}} \\
    Cirrus          & \texttt{\detokenize{r'\.cirrus\.yml\$'}} \\
    Wercker         & \texttt{\detokenize{r'wercker\.yaml\$'}} \\
    Semaphore       & \texttt{\detokenize{r'semaphore\.yml\$'}} \\
    Codemagic       & \texttt{\detokenize{r'codemagic\.yaml\$'}} \\
\hline
\end{tabular}
}
\vspace{-6pt}
\end{table}

\vspace{2pt}
\noindent\textbf{Step 5: Cloning final repositories.}
Following prior approaches, we shallow-clone each qualifying repository at its default branch and extract analysis artifacts. In addition to collecting CI configuration files (hereafter, ``YML files''; \texttt{.yml} or \texttt{.yaml}), we also resolve workflow-referenced artifacts that workflows call indirectly. Concretely, when a workflow refers to a repository script (e.g., a \texttt{.sh} file) or a local GitHub Action (e.g., \texttt{uses: ./...}), we locate the referenced file(s) and store them using their referenced paths. This allows us to analyze each workflow together with the scripts/local actions it calls when detecting instrumentation-testing signals. We also extract Gradle build scripts (\texttt{*.gradle} and \texttt{*.gradle.kts}) for analysis. This completes the pipeline with 4{,}518 successful clones, yielding 12{,}677 CI configuration files and 21{,}339 Gradle build scripts.

\subsection{Detecting Instrumentation Testing in CI Pipeline}
\label{sec:ci-detection}

This section explains how we detect repositories whose CI workflows execute Android instrumentation tests. Our focus is on CI-layer evidence: workflow YML files and the commands they execute, including inline commands, referenced scripts, and local actions.

\subsubsection{Detection criteria}
Executing Android instrumentation tests in CI typically requires (i) an Android execution environment (e.g., an emulator/device runtime or device-lab service) and (ii) an explicit test invocation (e.g., Gradle Android test tasks or Android Debug Bridge (ADB) instrumentation commands)~\cite{android-instrumented-tests,android_emulator_docs,android-test-cli}. We treat CI instrumentation adoption as a repository-level property: a repository is an adopter if any of its CI workflows contains explicit evidence of Android instrumentation-test execution-environment or test invocation.

\vspace{3pt}
\noindent\textbf{(A) Android execution-environment signals.}
Examples include installing/using the Android SDK, configuring an emulator/Android Virtual Device (AVD), launching an emulator~\cite{android_emulator_docs}, or integrating with a device-lab service (e.g., vendor CLI wrappers such as Firebase Test Lab)~\cite{google_github_actions_firebase_test_lab}. We also include Android-specific steps such as using \texttt{adb}~\cite{android-test-cli}.

\vspace{3pt}
\noindent\textbf{(B) Instrumentation-test invocation signals.}
We look for workflow steps that clearly run instrumentation tests. In practice, this often appears as a Gradle command that triggers device-based tests (e.g., \texttt{connectedAndroidTest} or \texttt{deviceCheck})~\cite{android-instrumented-tests} or as an explicit Android test command (e.g., an \texttt{adb instrumentation} run)~\cite{android-test-cli}. We also count device-lab or vendor commands when they clearly indicate that instrumentation tests are executed~\cite{google_github_actions_firebase_test_lab}.

\paragraph*{Scope note}
We treat execution environment and test invocation as independent components of CI-based instrumentation testing and use both to detect CI instrumentation testing.
However, \emph{execution environments} are central to reproducibility and CI reliability (e.g., device/runtime configuration and drift)~\cite{android_emulator_docs,android-instrumented-tests,Lam2020Lifecycle},
and exhibit greater cross-project configuration diversity (wrappers vs.\ custom setup vs.\ GMD vs.\ device labs)~\cite{android_emulator_docs,gradle_gmd_docs,google_github_actions_firebase_test_lab}.
By contrast, invocation is often standardized once established and may be less directly observable when mediated by scripts/actions or CI defaults~\cite{github_reuse_workflows,github_workflow_syntax,travis_android_default_test_commands}.
Accordingly, we focus our analyses on \emph{execution environments} and their changes over time.

\paragraph*{Flutter repositories (Android-scoped CI detection).}
Our CI-layer detection is \emph{Android-scoped}. For Flutter-related repositories, we only count CI evidence when it clearly targets Android execution (e.g., emulator/device provisioning, Android SDK/\texttt{adb} usage, or Android test invocation such as \texttt{connectedAndroidTest} or \texttt{adb shell am instrument})~\cite{android_emulator_docs,android-test-cli,android-instrumented-tests}. Thus, Flutter-related results should be interpreted as \emph{Android-targeted} CI instrumentation execution rather than overall Flutter testing in CI.

\subsubsection{Gradle Managed Devices (GMD) in build scripts}
\label{sec:gmd-build}

Some projects run instrumentation tests through Gradle Managed Devices (GMD), where the emulator/device setup is defined in Gradle rather than in the CI workflow~\cite{gradle_gmd_docs}. To capture this, we scan Gradle build scripts (\texttt{*.gradle} and \texttt{*.gradle.kts}) for GMD configuration blocks and for Gradle tasks associated with running tests on managed devices. When a CI workflow invokes such tasks, we use the Gradle-side GMD configuration as supporting evidence that instrumentation tests are intended to run in CI.

\subsection{Validation by Manual Review}
\label{sec:validation}

To assess the accuracy of our CI instrumentation-test detection, we conducted a manual validation on a statistically grounded random sample of the $N = 4{,}518$ CI-enabled repositories.

\vspace{3pt}
\noindent\textbf{Sample size and sampling procedure.}
We used a standard finite-population sample-size calculation for estimating a proportion with 90\% confidence, $\pm$10\% margin of error, and a conservative population proportion of 0.5 (worst case). This yields a required sample size $n = 68$ repositories. We draw a simple random sample without replacement using a fixed random seed to ensure reproducibility.

\vspace{3pt}
\noindent\textbf{Manual review protocol.}
For each sampled repository, we manually inspected its CI workflows and the workflow-expanded artifacts used by our detector (inline commands, workflow-referenced scripts, and local actions). We then determined whether the CI pipeline contains explicit evidence of Android instrumentation-testing in CI, operationalized as a concrete execution-environment signal and/or a concrete test-invocation signal in an Android context, matching the definition used in our automated pipeline.

\vspace{3pt}
\noindent\textbf{Accuracy.}
Across the 68 reviewed repositories, our automated classification matches the manual judgement in 66 cases, yielding an overall agreement/accuracy of $66/68 \approx 97.1\%$. Only two cases were mismatched. These mismatches were due to boundary situations where CI evidence was atypical or ambiguous (e.g., workflows that configure an Android execution environment but do not clearly execute Android instrumentation tests, or cases where test execution is triggered indirectly and is not explicit in the workflow text). Overall, this manual validation indicates that our CI instrumentation-test detection has high accuracy on the studied population.

\subsection{Analysis Procedures of RQs}
\label{sec:rq-methods}

This section describes the analysis procedures used in each research question (RQ1--RQ3), including the units of analysis and operationalization steps used in our empirical study.

\vspace{4pt}
\noindent\textbf{RQ1: What are the current adoption practices for instrumentation testing in CI?}
\label{sec:rq1-method}
To characterize current instrumentation-testing adoption in Android CI, we examine how often it is used and how execution environments are configured across open-source projects.
For repositories identified as CI instrumentation-test adopters (Section~\ref{sec:ci-detection}), we infer which execution-environment styles appear in their CI workflows by analyzing inline commands, workflow-called scripts, local actions, and Gradle signals for GMD (Section~\ref{sec:gmd-build})~\cite{gradle_gmd_docs}.

\vspace{3pt}
\noindent\textbf{Style labels.}
We use six labels:
\begin{itemize}
  \item \textbf{Community}: community-maintained wrappers/actions that encapsulate emulator setup/startup on CI runners~\cite{android_emulator_docs}.
  \item \textbf{Custom}: project-specific scripting for emulator/device setup and execution.
  \item \textbf{GMD}: Gradle Managed Devices-based execution where device definitions are declared in Gradle~\cite{gradle_gmd_docs}.
  \item \textbf{Third Party}: third-party device-lab/device-farm services (e.g., Firebase Test Lab)~\cite{google_github_actions_firebase_test_lab}.
  \item \textbf{Real Device}: execution on physical devices explicitly indicated in repository artifacts.
  \item \textbf{Unclassified}: CI clearly runs instrumentation tests, but the emulator/device setup cannot be mapped to our main styles because the environment is not explicitly visible in the repository artifacts we analyze (e.g., remote marketplace actions treated as black boxes, non-standard wrappers, or implicit/externally configured device execution)~\cite{github_actions_building_blocks,github_self_hosted_runners}.
\end{itemize}

\noindent\textbf{Any-use vs. Single-style counting.}
A repository can use multiple styles across different workflows; therefore, we report:
(i) \emph{Any-use} prevalence using multi-label counting (each style counted at most once per repository), and
(ii) \emph{Single-style} comparisons restricted to repositories that use exactly one style.
Repositories using multiple styles are reported descriptively and excluded from statistical comparisons.

\vspace{5pt}
\noindent\textbf{RQ2: How do execution environments evolve over time on the default branch?}
\label{sec:rq2-method}
Given that a project’s current CI setup reflects only its latest state, we study how execution-environment styles evolve over time by focusing on repositories with at least one detected execution-environment episode traceable on the default branch up to the cut-off date. For this analysis, we retrieve sufficient default-branch commit history (by deepening/fetching repository history as needed) to traverse execution-environment evidence up to the cut-off date for the included adopter repositories.

\vspace{3pt}
\noindent\textbf{Input set.}
From the $N=481$ adopter repositories identified by our CI instrumentation detector, we include repositories with a detectable and traceable history for the execution-environment evidence signals on the default branch (yielding $N=434$ repositories used for evolution analysis).

\vspace{3pt}
\noindent\textbf{Episode definition.}
We model evolution as \emph{change episodes}: maximal contiguous time intervals over the default-branch history during which the execution-environment style set remains unchanged. This design follows prior work on CI pipeline evolution and CI-tool adoption/migration~\cite{zampetti2021cicd,mazrae2023cicdtools}.

\vspace{3pt}
\noindent\textbf{Timeline extraction.}
For each repository, we traverse the default branch (mainline) history up to the cut-off date and map each snapshot to a canonical execution-environment style set using the same taxonomy and evidence functions as RQ1. We use a first-parent traversal to avoid inflating changes from short-lived side branches and to approximate merged history.

\vspace{3pt}
\noindent\textbf{Retention rule (noise reduction).}
To reduce noise from transient edits, we retain an episode only if it (i) persists for at least 14 days, or (ii) is still active at the cut-off date~\cite{Lam2020Lifecycle,Shimagaki2016Reverts}. Episodes are labeled by the single style when exactly one style is present, and labeled as \textbf{Mixed} when multiple styles appear.

\vspace{3pt}
\noindent\textbf{State-machine summarization.}
We summarize episode-to-episode transitions as a state machine over the observed style states, consistent with event-log/process-mining approaches that derive compact transition models from temporal traces~\cite{aalst2016processmining}.

\vspace{3pt}
\noindent\textbf{RQ3: What drives projects to change instrumentation-test environments, and how does CI performance vary by style?}
\label{sec:rq3-method}
Given that configuration changes alone do not explain developer intent, we examine (i) text-mined intention signals around episode boundaries and (ii) CI run telemetry that characterizes operational performance under different execution-environment styles.

\paragraph*{Motivation (intention) extraction from boundary context}
To interpret why repositories change execution environments, we analyze boundary commits associated with environment adds/removals/transitions using commit, pull request (PR) and issue text and categorize them into a fixed set of developer-facing motivation labels (Table~\ref{tab:intention-labels}). For each boundary commit, we build a weighted boundary document by concatenating available context fields (commit subject/body; PR title/body and discussion; issue title/body and discussion). We follow a Subcat-style dictionary-based approach for intention classification~\cite{mauczka2012subcat}, using term frequency-inverse document frequency (TF--IDF) weighting~\cite{salton1988tfidf} with term specificity derived from inverse document frequency (IDF)~\cite{sparckjones1972idf} (background:~\cite{manning2008iir}). 
We mine candidate $n$-grams (up to 3-grams) to refine dictionaries and rank candidates using an IDF-like specificity factor~\cite{sparckjones1972idf,salton1988tfidf,papagiannopoulou2019keyphrase}. We allow up to three labels per boundary, apply rule-based constraints to suppress implausible labels, and assign an automatic confidence level (HIGH/MEDIUM/LOW), following prior dictionary-based validation practice~\cite{mauczka2012subcat}.

\begin{table*}[t]
\caption{Motivation labels (short description + representative example commit message).}
\label{tab:intention-labels}
\vspace{-3pt}
\renewcommand{\arraystretch}{1.05}
\centering
\begin{tabularx}{\textwidth}{
  @{}
  >{\raggedright\arraybackslash}p{0.04\textwidth}
  >{\raggedright\arraybackslash}p{0.26\textwidth}
  >{\raggedright\arraybackslash}p{0.22\textwidth}
  >{\raggedright\arraybackslash}X
  @{}
}
\toprule
\textbf{No.} & \textbf{Label} & \textbf{Description} & \textbf{Example commit message} \\
\midrule
1 & Address performance or stability issues & Fix flaky/slow emulator tests. &
\texttt{test(app): stablise flaky tests after removing retry test rule (\#782)} \\

2 & Expand test scope or capabilities & Add/expand instrumentation coverage. &
\texttt{Automated screenshot testing} \\

3 & Introduce / strengthen CI-backed tests & Enable tests in CI (gates/coverage). &
\texttt{[Build] Code coverage report and run android test with emulator (\#38)} \\

4 & Clean up or simplify CI / environment configuration & Simplify CI + emulator setup. &
\texttt{ci: remove .travis.yml} \\

5 & Migrate or modernize CI infrastructure & Move CI to newer tooling. &
\texttt{ci(github): migrate to GitHub actions} \\

6 & Automate or integrate release workflows & Automate build/sign/release. &
\texttt{feat(qa-build): update pipeline for generating ios and android builds (\#17601)} \\
\bottomrule
\end{tabularx}
\vspace{-8pt}
\end{table*}

\paragraph*{CI run telemetry and operational metrics}
To complement textual intent, we derive operational profiles for execution-environment styles using GitHub Actions run telemetry. We retain repositories for which we can observe instrumentation-test runs on GitHub Actions and extract run-level outcomes (conclusion), runtime duration, and rerun attempts. Where available, we also extract step-level timestamps to attribute time to environment setup versus test execution and supporting steps, and we report descriptive breakdowns to support interpretation~\cite{bouzenia2024actionsresources,zheng2025actionsfail}.

\subsection{Statistical Analysis}
\label{sec:stats}

Because this is an observational study, we report associations only and do not infer causality.

\vspace{3pt}
\noindent\textbf{Categorical associations.}
For associations between categorical variables (e.g., adoption vs.\ team-size group), we use the $\chi^2$ test of independence~\cite{pearson1992,agresti2013}. For $2\times 2$ contingency tables, we report $\phi$ as the effect size; for larger tables, we report Cram\'er's $V$~\cite{agresti2013}. Following common benchmarks for effect sizes~\cite{cohen1988}, we interpret $\phi$ (and, heuristically, Cram\'er's $V$) as follows: $|e| < 0.10$ is \emph{negligible}, $0.10 \le |e| < 0.30$ is \emph{small}, $0.30 \le |e| < 0.50$ is \emph{medium}, and $|e| \ge 0.50$ is \emph{large}.

\vspace{3pt}
\noindent\textbf{Multiple comparisons.}
When performing families of related hypothesis tests (e.g., the style-specific intention skews in Table~VI), we control the false discovery rate using the Benjamini--Hochberg procedure and report adjusted $p$-values as \emph{pFDR}~\cite{benjamini1995fdr}.

\vspace{3pt}
\noindent\textbf{Proportion differences.}
When comparing two proportions, we report Cohen's $h$~\cite{cohen1988}. We interpret $|h|<0.20$ as \emph{negligible}, $0.20 \le |h| < 0.50$ as \emph{small}, $0.50 \le |h| < 0.80$ as \emph{medium}, and $|h|\ge 0.80$ as \emph{large}~\cite{cohen1988}.

\vspace{3pt}
\noindent\textbf{Heavy-tailed numeric metrics.}
For repository-level numeric metrics that are heavy-tailed (e.g., commits, pull requests, issues), we use one-sided Mann--Whitney $U$ tests~\cite{mann1947}. We report Cliff's $\delta$ as an ordinal effect size~\cite{cliff1993} and interpret its magnitude using commonly cited thresholds: $|\delta|<0.147$ \emph{negligible}, $0.147 \le |\delta| < 0.33$ \emph{small}, $0.33 \le |\delta| < 0.474$ \emph{medium}, and $|\delta|\ge 0.474$ \emph{large}. The sign of $\delta$ indicates direction (positive when the first group tends to have larger values).~\cite{romano2006}

\noindent\textbf{Implementation.}
All statistical tests are implemented in Python using SciPy~\cite{virtanen2020}.

\section{Empirical Results}
\label{sec:results}

This section reports empirical findings for RQ1--RQ3 based on the dataset and analysis procedures described in Section~\ref{sec:methodology}. For each research question, we summarize the key observations and supporting evidence.

\subsection{RQ1 Results.}
\label{sec:RQ1}

\vspace{3pt}
\noindent\textbf{Observation 1.1: Instrumentation testing in CI is uncommon among CI-enabled repositories.}
Across $N=4{,}518$ CI-enabled repositories, 481 repositories (10.6\%) execute Android instrumentation tests in CI workflows under our repository-level adoption definition (Section~\ref{sec:ci-detection}).  This observation motivates the stakeholder implications in Section~\ref{sec:implications}.

\vspace{2pt}
\noindent\textbf{Observation 1.2: Adoption differs by primary language.}
Using GitHub’s primary language, Kotlin+Java repositories execute instrumentation tests in CI at 13.4\% (386/2{,}886),
whereas \emph{C-family}-primary repositories (C/C++/Objective-C variants) do so at only 2.7\% (10/365).
This gap is statistically significant ($\chi^2{=}34.26$, $p{=}4.82\times10^{-9}$) and has a non-trivial proportion difference (Cohen’s $h{=}0.42$).

\vspace{2pt}
\noindent\textbf{Observation 1.3: Most adopters use a single environment style, and Community environments dominate.}
Most adopters use exactly one environment style (460/481 = 95.6\%); mixed-style usage is uncommon (21/481 = 4.4\%).
Among all adopters ($N{=}481$), execution environments are dominated by \emph{Community} wrappers/actions (251; 52.2\% any-use) and \emph{Custom} scripting (150; 31.2\%), while \emph{GMD} (16; 3.3\%), \emph{Third Party} device labs (32; 6.7\%), and \emph{Real Device} execution (2; 0.4\%) are comparatively rare; 51 repositories (10.6\%) are \emph{Unclassified}.
Table~\ref{tab:rq1-env-styles} summarizes these counts and highlights that dominance is likely driven by single-style usage rather than frequent configuration mixing.

\begin{table}[ht]
\centering
\caption{Distribution of execution-environment styles among CI instrumentation-test adopters ($N{=}481$), reported as \emph{Any-use} (a repo uses the style in at least one workflow; multi-label) and \emph{Single-style} (repos using exactly one style).}

\vspace{-3pt}
\label{tab:rq1-env-styles}
\small
\setlength{\tabcolsep}{5pt}
\begin{tabular}{@{}lrr@{}}
\toprule
\textbf{Style} & \textbf{Any-use} & \textbf{Single-style} \\
\midrule
Community     & 251 (52.2\%) & 232 (48.2\%) \\
Custom        & 150 (31.2\%) & 141 (29.3\%) \\
GMD           &  16 ( 3.3\%) &   7 ( 1.5\%) \\
Third Party   &  32 ( 6.7\%) &  27 ( 5.6\%) \\
Real Device   &   2 ( 0.4\%) &   2 ( 0.4\%) \\
Unclassified  &  51 (10.6\%) &  51 (10.6\%) \\
\midrule
Mixed-style repos ($\ge$2 styles) & -- & 21 (4.4\%) \\
\bottomrule
\end{tabular}
\vspace{-4pt}
\end{table}

\begin{figure}[t]
  \centering
  \includegraphics[width=\linewidth]{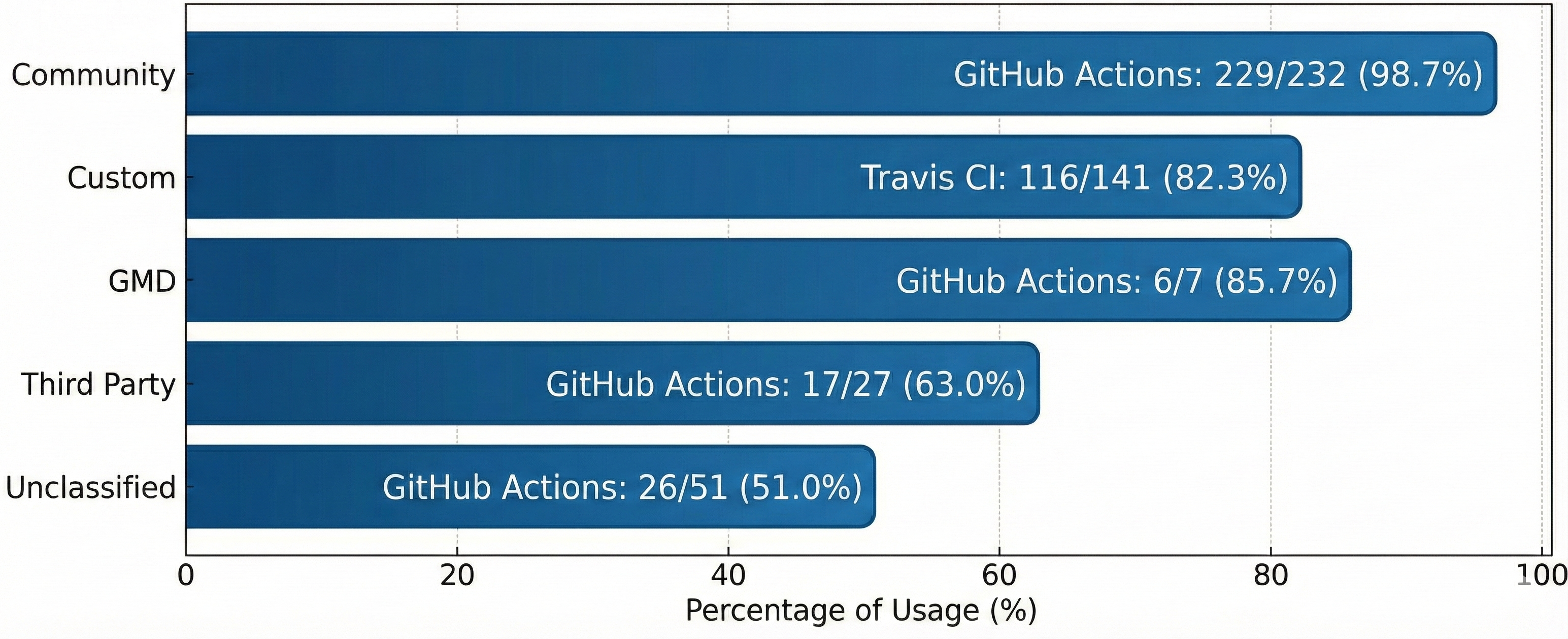}
  \vspace{-10pt}
  \caption{Dominant CI service provider by execution-environment style among single-style adopters (excluding Real Device); repositories may use multiple CI services.}
  \label{fig:rq1-style-ci-rate}
  \vspace{-14pt}
\end{figure}

\vspace{2pt}
\noindent\textbf{Observation 1.4: Execution style is strongly coupled with CI provider usage, and style strata differ sharply in activity/age.}
We stratify by execution style to compare CI-provider usage (Figure~\ref{fig:rq1-style-ci-rate}), restricting to \emph{single-style} adopters (excluding \emph{Mixed} and \emph{Real Device} as rare cases in Observation~1.3). We include \emph{Unclassified} as a separate \emph{style-not-observed} category (i.e., confirmed execution but missing environment evidence). 
Figure~\ref{fig:rq1-style-ci-rate} shows strong provider concentration by style: \emph{Community} aligns almost exclusively with GitHub Actions, \emph{Custom} aligns primarily with Travis CI, and \emph{GMD} and \emph{Third Party} also skew toward GitHub Actions but less strongly; \emph{Unclassified} remains provider-skewed as well. 
These associations are statistically decisive when tested as provider \emph{presence} (binary) by style, with large effects for GitHub Actions ($\chi^2{=}289.31$, $p{=}2.19\times10^{-61}$, Cramér’s $V{=}0.795$) and Travis CI ($\chi^2{=}270.27$, $p{=}2.79\times10^{-57}$, $V{=}0.768$).

Beyond CI coupling, the dominant styles also differ sharply in repository characteristics: \emph{Community}-style repositories are more active than \emph{Custom}-style repositories, since it is statistically verified that they have significantly more pull requests and commits (one-sided Mann--Whitney $U$: pull requests $p{=}1.77\times10^{-17}$, $\delta{=}0.520$; commits $p{=}8.66\times10^{-13}$, $\delta{=}0.436$), while \emph{Custom}-style repositories tend to be older (repo age $p{=}1.88\times10^{-19}$, $\delta=-0.552$).
These results suggest that execution-environment choice (or lack of observable evidence for it) co-occurs with broader CI platform choices and with recognizable differences in project maturity and activity.

\vspace{2pt}
\noindent\textbf{Observation 1.5: When observable, test invocation is overwhelmingly Gradle-based.}
Test invocation is explicitly observable in 360/481 adopters (74.8\%). When observable, it is overwhelmingly Gradle-based (317/360 = 88.1\% any-use), with fewer repositories using provider CLIs (41/360 = 11.4\%) or direct ADB/shell commands (7/360 = 1.9\%). Mixed invocation mechanisms are rare (5/481 = 1.0\% of adopters).

\vspace{3pt}
\begin{rqbox}
\textbf{RQ1 Summary:} CI instrumentation execution is rare and ecosystem-dependent, with much higher adoption in Kotlin/Java than C/C++. Among adopters, \emph{Community} and \emph{Custom} dominate with little mixing; environment choice aligns strongly with CI providers and differentiates project profiles (Community: more active; Custom: older). Observable invocations are overwhelmingly Gradle-based.
\end{rqbox}

\subsection{RQ2 Results.}
\label{sec:RQ2}

\begin{figure}[ht]
  \centering
  \includegraphics[width=.9\linewidth]{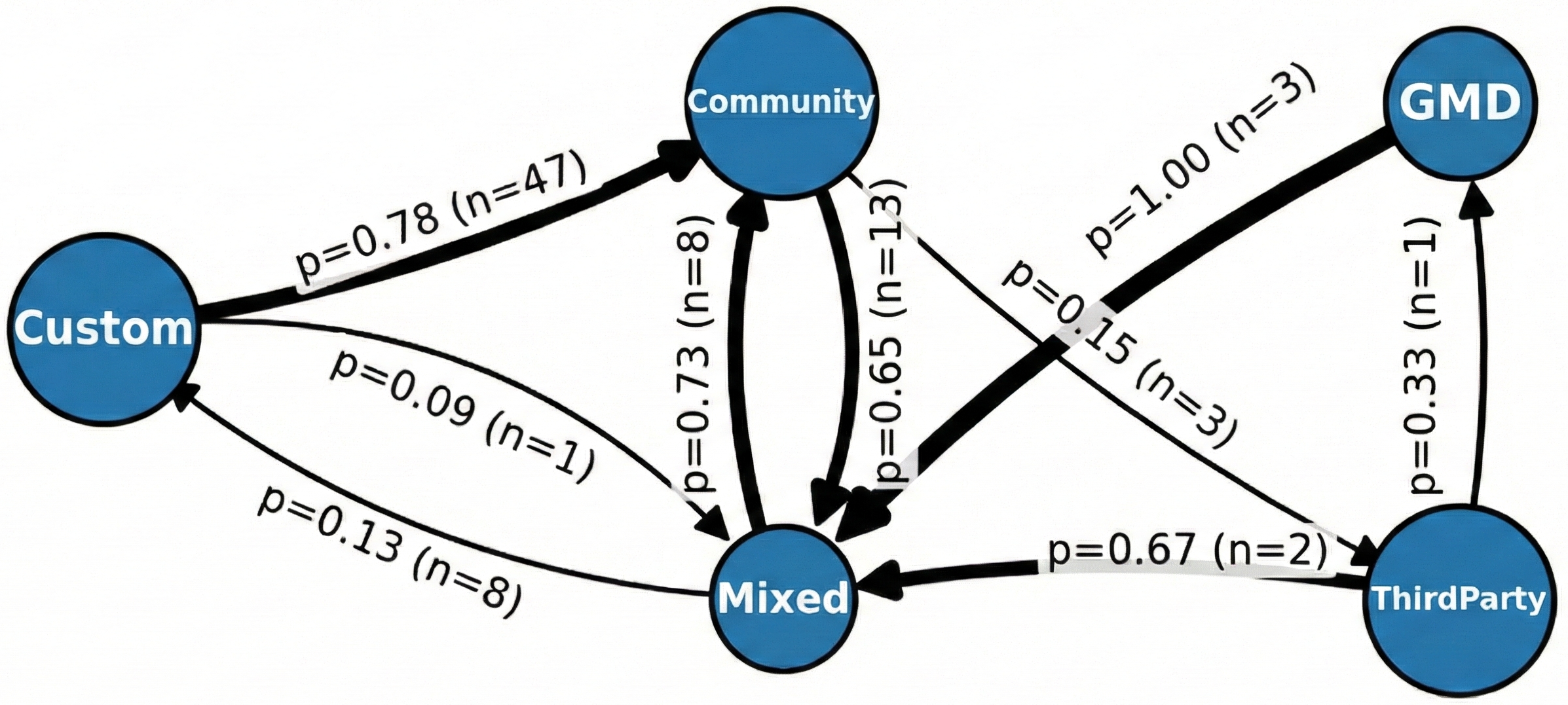}
  \vspace{-2pt}
  \caption{Episode-to-episode state machine of execution-environment evolution (only the top-2 outgoing transitions per state are shown for readability). Edge labels: transition probability ($p$) and count ($n$).}
  \label{fig:rq2-state-machine}
  \vspace{-10pt}
\end{figure}

\noindent\textbf{Observation 2.1: Custom-first histories converge to Community.}
Repositories tend to converge toward \emph{Community}: although \emph{Custom} is the most common entry state, \emph{Community}
becomes the dominant current state by the cutoff (Table~\ref{tab:rq2-entry-current-compact}). This shift matches the state machine:
the dominant migration is \emph{Custom}$\rightarrow$\emph{Community} (47/101 transitions), and departures from \emph{Community}
(often via \emph{Mixed}) frequently resolve back to \emph{Community} (Figure~\ref{fig:rq2-state-machine}). This observation motivates the stakeholder implications in Section~\ref{sec:implications}.

\begin{table}[ht]
  \centering
  \small
  \caption{RQ2 entry vs.\ current distribution (headline states; $N{=}434$).}
  \vspace{-3pt}
  \label{tab:rq2-entry-current-compact}
  \begin{tabular}{lcc}
    \toprule
    \textbf{State} & \textbf{Entry} & \textbf{Current} \\
    \midrule
    Community & 192 (44.2\%) & 227 (52.3\%) \\
    Custom    & 204 (47.0\%) & 147 (33.9\%) \\
    Other (Mixed+GMD+ThirdParty) & 38 ( 8.8\%) & 60 (13.8\%) \\
    \bottomrule
  \end{tabular}
  \vspace{-5pt}
\end{table}

\begin{figure}[ht]
  \centering
  \small
  \includegraphics[width=\linewidth]{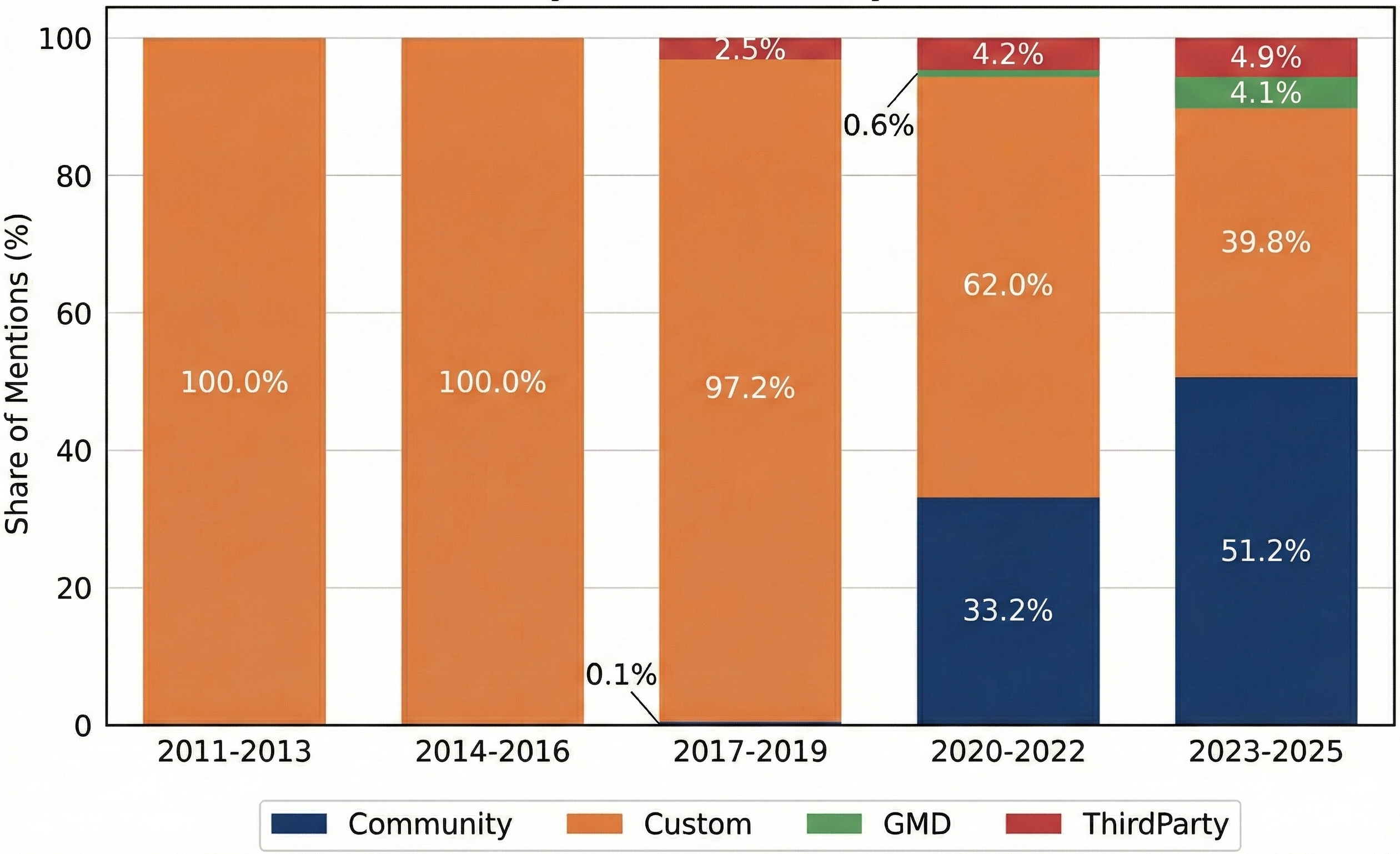}
  \caption{Time-sliced execution-style presence aggregated into 3-year buckets (style shares by bucket).}
  \label{fig:rq2_stackedbar}
  \vspace{-5pt}
\end{figure}

\vspace{2pt}
\noindent\textbf{Observation 2.2: Environment changes are rare, and migrations cluster in 2020--2022.}
Environment changes are uncommon: 354/434 repositories (81.6\%) retain a single episode (no transition), and only 80/434 (18.4\%) exhibit at least one transition (up to four episodes per repository). When transitions occur, they are temporally concentrated: 36/47 \emph{Custom}$\rightarrow$\emph{Community} transitions (76.6\%) begin in 2020--2022, suggesting a short migration wave rather than steady change. The same shift appears in the 3-year buckets, where early periods are dominated by \emph{Custom} and \emph{Community} rises sharply in 2020--2022 (Figure~\ref{fig:rq2_stackedbar}). This observation motivates the stakeholder implications in Section~\ref{sec:implications}.

\emph{Note:} Buckets are time-sliced style presence, while “entry” is repository-specific; many repositories first appear after 2016,
so their entry state can be \emph{Community} even if early-year buckets are entirely \emph{Custom}.

\vspace{2pt}
\noindent\textbf{Observation 2.3: Mixed is usually a transitional waypoint, not a destination.}
\emph{Mixed} episodes (two or more concurrent styles) are rare overall but appear disproportionately around migrations. The state machine
is consistent with \emph{Mixed} functioning as a waypoint: \emph{Community}$\rightarrow$\emph{Mixed} occurs 13 times and
\emph{Mixed}$\rightarrow$\emph{Community} occurs 8 times, while few repositories remain in \emph{Mixed} at the cutoff (26/434, 6.0\%).
Mixed episodes most commonly represent \emph{Community+Custom} (17/41, 41.5\%) or \emph{Community+GMD} (13/41, 31.7\%), indicating that
overlap typically occurs during adoption of community (and occasionally while integrating GMD).

\vspace{3pt}
\begin{rqbox}
\textbf{RQ2 Summary:} Execution environment changes are infrequent and predominantly move from \emph{Custom} to \emph{Community}, reflecting the dominant \emph{Custom}$\rightarrow$\emph{Community} migration. Most projects never migrate, changes cluster in 2020--2022, and \emph{Mixed} typically appears only as a brief transition state.
\end{rqbox}

\vspace{3pt}
\subsection{RQ3 Results.}
\label{sec:rq3}

\vspace{2pt}
\noindent\textbf{Observation 3.1: Adoption (add a style) motivations are dominated by test enablement and scope expansion.}
Across adoption (Added) commits with at least one \emph{selected} intention label (multi-labeling possible), we observe that detected intentions concentrate primarily on (i) expanding test scope/capabilities and (ii) introducing/strengthening CI-backed tests (together 57.6\% of detections; $k=167/290$). The remaining detections are distributed across stability/performance work, CI/environment configuration cleanup, CI modernization, and release automation (Table~\ref{tab:obs_adoption_intent_breakdown}).

\begin{table}[ht]
\centering
\renewcommand{\arraystretch}{1.05}
\caption{motivation breakdown for adoption (Added) commits with non-empty \texttt{intent\_label\_str} ($n=261$ labeled commits). \emph{Note:} multi-labeling means total $k$ (intentions) can exceed the number of labeled commits.}
\vspace{-3pt}
\resizebox{\linewidth}{!}{
\begin{tabular}{p{0.70\columnwidth} r r}
    \hline
    \textbf{Detected intention (Added, selected labels)} & \textbf{$k$} & \textbf{(\%)} \\
    \hline
    Expand test scope or capabilities             & 85  & 29.3 \\
    Introduce / strengthen CI-backed tests        & 82  & 28.3 \\
    Address performance or stability issues       & 41  & 14.1 \\
    Clean up or simplify CI / env.\ configuration & 30  & 10.3 \\
    Automate or integrate release workflows       & 24  & 8.3  \\
    Migrate or modernise CI infrastructure        & 28  & 9.7  \\
    \hline
    \textbf{Total}                                & 290 & 100.0 \\
    \hline
\vspace{-5pt}
\end{tabular}
}
\label{tab:obs_adoption_intent_breakdown}
\vspace{-8pt}
\end{table}

\vspace{2pt}
\noindent\textbf{Observation 3.2: Adoption motivation differs by environment style (evidence-backed detections).}
Using adoption (Added) boundary events with evidence-backed intention detections (non-zero text match), the intention distribution differs significantly by environment style ($p=7.09\times10^{-6}$, $V=0.218$). The most pronounced style-specific skews are summarized in Table~\ref{tab:obs_adoption_style_skews}: GMD adoption shows a stronger cleanup/simplification signal, ThirdParty adoption emphasizes expanding test scope/capabilities, and Community adoption is more frequently associated with CI migration/modernization.

\begin{table*}[ht]
\centering
\caption{Style-specific intention skews for adoption (Added) boundary events with evidence-backed detections. \emph{Note:} $k$ counts detected intentions (not commits); $n$ is the total number of intentions within each group.}
\vspace{-3pt}
\resizebox{\linewidth}{!}{
\renewcommand{\arraystretch}{1.15}
    \begin{tabular}{l l | l l | l}
    \hline
    \textbf{Style} & \textbf{Intention} &
    \textbf{Style} $k/n$ (\%) & \textbf{Others} $k/n$ (\%) & \textbf{Stats} \\
    \hline
    GMD &
    Clean up or simplify CI / environment configuration &
    $7/26$ (26.9) & $32/333$ (9.6) &
    $p_{\mathrm{FDR}}=1.44\times10^{-2};\ \phi=0.144$ \\
    Third Party &
    Expand test scope/capabilities &
    $17/34$ (50.0) & $84/325$ (25.8) &
    $p_{\mathrm{FDR}}=6.95\times10^{-3};\ \phi=0.157$ \\
    Community &
    Migrate/modernise CI infrastructure &
    $37/220$ (16.8) & $5/139$ (3.6) &
    $p_{\mathrm{FDR}}=2.34\times10^{-4};\ \phi=0.200$ \\
    \hline
    \end{tabular}
}
\label{tab:obs_adoption_style_skews}
\vspace{-7pt}
\end{table*}

\vspace{2pt}
\noindent\textbf{Observation 3.3: Third Party styles in workflows are predominantly scheduled, while Community/Custom are mostly push- and PR-driven.}
Workflow triggers differ sharply by environment style (computed over all runs in each style: \texttt{Third Party} $n{=}3{,}100$, \texttt{Community} $n{=}26{,}054$, \texttt{Custom} $n{=}363$). \texttt{Third Party} runs are primarily \emph{scheduled} (\textbf{66.97\%}, i.e., 2{,}076/3{,}100), with fewer triggered by direct code pushes (\textbf{24.06\%}, 746/3{,}100). In contrast, \texttt{Community} runs are mostly push-driven (\textbf{79.30\%}, 20{,}662/26{,}054) with a smaller scheduled share (\textbf{15.08\%}, 3{,}928/26{,}054). \texttt{Custom} is also largely push-driven (\textbf{81.82\%}, 297/363) but shows a higher proportion of pull-request-triggered runs (\textbf{17.08\%}, 62/363). This indicates that \texttt{Third Party} environments are often exercised in periodic (e.g., nightly/regression) workflows rather than interactive change-driven workflows, which is important context when interpreting performance and reliability differences across styles.

\vspace{2pt}
\noindent\textbf{Observation 3.4: Third Party success rate is dominated by startup failures (vs.\ Community stability).}
success rate differs substantially by environment style (\autoref{tab:rq3-availability}). Over \texttt{Community} runs ($n{=}26{,}054$), \textbf{65.25\%} succeed (17{,}000/26{,}054), with \textbf{23.42\%} failing (6{,}103/26{,}054) and \textbf{10.19\%} cancelled (2{,}656/26{,}054). In contrast, \texttt{Third Party} succeeds only \textbf{32.23\%} of the time (999/3{,}100) and is heavily impacted by environment startup failures (\textbf{32.26\%}, 1{,}000/3{,}100), a mode that is essentially absent in \texttt{Community} (\textbf{0.01\%}, 3/26{,}054). Even after excluding startup failures, \texttt{Third Party}'s success rate reaches only \textbf{47.57\%} (999/2{,}100), still well below \texttt{Community} (\textbf{65.26\%}, 17{,}000/26{,}051). Consistent with the interpretation of \emph{startup} failures, none of the runs labeled \texttt{startup\_failure} exhibit any job/step execution trace in our step-level breakdown, indicating that these failures occur before workflow jobs begin. Since GMD is known to be uncommon overall (per our earlier RQs) and this dataset contains no GMD-only runs, we report the only GMD evidence available here: the mixed \texttt{Community+GMD} group ($n{=}51$). This observation motivates the stakeholder implications in Section~\ref{sec:implications}.

\begin{table}[ht]
\centering
\caption{success rate by environment style. Percentages are computed over the run counts. ``Community+GMD'' is a mixed-style label; this dataset contains no GMD-only runs.}
\vspace{-3pt}
\renewcommand{\arraystretch}{1.05}
\resizebox{\linewidth}{!}{
    \begin{tabular}{lrrrrr}
    \hline
    Style & Runs & Success & Failure & Cancelled & Startup \\
    \hline
    Community & 26{,}054 & 65.3\% & 23.4\% & 10.2\% & 0.0\% \\
    Third Party & 3{,}100 & 32.2\% & 34.5\% & 0.9\% & 32.3\% \\
    Custom & 363 & 70.3\% & 27.6\% & 2.2\% & 0.0\% \\
    Community+GMD & 51 & 88.2\% & 11.8\% & 0.0\% & 0.0\% \\
    \hline
    \end{tabular}
}
\label{tab:rq3-availability}
\vspace{-2pt}
\end{table}

\vspace{2pt}
\noindent\textbf{Observation 3.5: Third Party runs are $\approx$1.8$x$ slower than Community and far more likely to exceed one hour.}
Runtime differs meaningfully by environment style (\autoref{tab:rq3-runtime-pressure}). Runtime summaries are computed over runs with non-null duration (e.g., \texttt{Community} $n{=}25{,}630$; \texttt{Third Party} $n{=}2{,}099$). \texttt{Community} has a median runtime of \textbf{12.4 minutes}, whereas \texttt{Third Party} has a median of \textbf{21.9 minutes} (about \textbf{1.8$x$} longer). This gap persists in the tail: the 95th percentile is \textbf{1.6 hours} for \texttt{Community} versus \textbf{2.1 hours} for \texttt{Third Party}. Long runs are also more frequent under \texttt{Third Party}, where \textbf{25.35\%} of runs exceed one hour (532/2{,}099) compared to \textbf{11.25\%} for \texttt{Community} (2{,}884/25{,}630). For \texttt{Community+GMD} (mixed, $n{=}51$), runtimes are shorter and tightly distributed (median \textbf{13.2 minutes}), but due to the small sample, we treat this as descriptive rather than conclusive. This observation motivates the stakeholder implications in Section~\ref{sec:implications}.

\begin{table}[ht]
\centering
\renewcommand{\arraystretch}{1.05}
\caption{
Runtime and queuing pressure by environment style. Runtime summaries are computed over runs with non-null duration
}
\vspace{-3pt}
\resizebox{\linewidth}{!}{
    \begin{tabular}{lrrrrr}
    \hline
    Style & Median & P95 & P99 & Queue$>$0 & Q$_{med}$ (nz) \\
    \hline
    Community & 12.4m & 1.6h & 7.8h & 4.00\% & 1.5h \\
    Third Party & 21.9m & 2.1h & 6.0h & 0.65\% & 7.0h \\
    Custom & 23.1m & 1.6h & 10.3h & 13.22\% & 3.0h \\
    Community+GMD & 13.2m & 21.2m & 24.4m & 0.00\% & -- \\
    \hline
    \end{tabular}
}
\label{tab:rq3-runtime-pressure}
\vspace{-7pt}
\end{table}

\vspace{3pt}
\noindent\textbf{Observation 3.6: Pressure mechanisms differ by style: Custom is rescheduling-heavy, while Third Party is runtime-heavy.}
\label{obs:rq3-pressure}
We quantify ``pressure'' using (i) queuing/rescheduling delay (waiting time; Queue$>$0 in \autoref{tab:rq3-runtime-pressure}) and (ii) long-run tail behavior. \texttt{Custom} shows the strongest rescheduling-driven pressure: \textbf{13.22\%} of runs have nonzero queuing delay (48/363), compared to \textbf{4.00\%} for \texttt{Community} (1{,}043/26{,}054) and \textbf{0.65\%} for \texttt{Third Party} (20/3{,}100); none for \texttt{Community+GMD} ($n{=}51$). In this dataset, nonzero waiting time occurs only on rerun attempts, indicating rerun rescheduling latency rather than general CI scheduler contention. By contrast, \texttt{Third Party} pressure is dominated by long-running jobs (Obs.~3.5). Where step traces are available (8{,}226 runs total; \texttt{Third Party} $n{=}711$, \texttt{Community} $n{=}7{,}421$, \texttt{Custom} $n{=}79$, \texttt{Community+GMD} $n{=}15$), provider steps account for essentially all extracted step time for \texttt{Third Party}. \texttt{Custom} allocates a larger share of extracted step time to local pipeline overhead (notably environment setup and artifact handling), consistent with its higher rescheduling burden. This observation motivates the stakeholder implications in Section~\ref{sec:implications}.

\vspace{2pt}
\begin{rqbox}
\textbf{RQ3 Summary:}
Repositories change testing environments mainly to expand instrumentation testing, but each environment serves a distinct CI role: \texttt{Community} is the most reliable and efficient option for push/PR gating, \texttt{Third Party} fits scheduled regression despite higher failure rates and cost, and \texttt{Custom} provides flexibility but incurs frequent reruns, reflecting workflow pressure rather than runtime limits.
\end{rqbox}

\section{Implications}
\label{sec:implications}
Our findings offer implications for practitioners, tool builders, CI and ecosystem providers, and researchers.

\vspace{2pt}
\noindent\textbf{Practitioners and tool builders} can leverage the taxonomy as a practical diagnostic and decision aid grounded in observed trade-offs. Since CI execution is uncommon (Obs.~1.1), enabling CI runs is a key first step for many “Gradle-only” projects. Among adopters, \texttt{Community} is the dominant style and a frequent migration target (Obs.~1.3, Obs.~2.1), \texttt{Custom} offers flexibility but incurs higher rerun and rescheduling pressure (Obs.~3.6), and \texttt{Third Party} is typically used for scheduled workflows (Obs.~3.3) while carrying higher startup-failure risk and longer runtimes (Obs.~3.4--3.5). This guidance helps teams choose environment styles aligned with their operational goals and workflow constraints.

\vspace{2pt}
\noindent\textbf{CI and ecosystem providers} can help reduce adoption friction (Obs.~1.1) and facilitate the common migration path toward community-style setups (Obs.~2.1--2.2) by providing robust, provider-agnostic templates and wrappers, improving onboarding and diagnostics for managed devices and device labs, and enhancing observability. In particular, better visibility can distinguish pre-job startup failures from actual test-execution failures (Obs.~3.4) and highlight where delays and pressure occur, whether from provider-side queuing, local setup overhead, or reruns (Obs.~3.5--3.6). This guidance can directly inform tooling and platform improvements to make CI instrumentation more reliable and accessible.

\vspace{2pt}
\noindent\textbf{Researchers} can leverage our execution-environment taxonomy and longitudinal episode model (Obs.~1.3, Obs.~2.1--2.3), boundary-intention signals (Obs.~3.1--3.2), and run/step metrics (Obs.~3.4--3.6) to investigate open questions in Android CI. These resources enable systematic study of why CI execution remains uncommon (Obs.~1.1), how environment choices affect reliability, runtime, and cost, and what drives migrations or persistent configurations. By providing structured, auditable evidence across CI and build layers, our dataset and models lay a foundation for deeper analyses of instrumentation-test adoption, operational trade-offs, and tool-supported improvements in CI practices.

\section{Threats to Validity}
\label{sec:threats_to_validty}
We outline threats to validity by \emph{construct}, \emph{internal}, and \emph{external} validity and note mitigations.

\vspace{2pt}
\noindent\textbf{Construct validity.}
CI configurations can be indirect (wrappers, helper scripts, reusable components), and some repositories keep \texttt{androidTest} scaffolding without running instrumentation tests in CI. To reduce false positives, we require explicit in-repo evidence of CI execution and assign execution-environment styles only when workflows or referenced code clearly indicate them; ambiguous cases remain \texttt{Unclassified}, likely lowering style counts. Given that \texttt{GMD} configuration may reside in Gradle rather than workflows, we supplement CI evidence with targeted build-layer mining for GMD signals.

In addition, RQ3 intent labels are derived from episode-boundary commit/PR/issue text and are intentionally conservative, as many boundaries lack context or use generic phrasing. In our dataset, 276/524 boundary commits (52.7\%) have at least one intent label; the rest are treated as \emph{unlabeled/unknown}, not evidence of absent intent. This partial coverage can bias intent distributions, so we use intents only for aggregate trends and report confidence levels for labeled cases.

RQ3 also uses workflow-run outcomes and step-level timing as proxies for reliability and cost. Outcomes can conflate causes (test flakiness, infrastructure issues, cancellations), and step-time attribution is heuristic; we use step breakdowns to support interpretation rather than as primary evidence. The ``waiting'' metric mainly reflects rerun rescheduling latency, which we interpret together with rerun attempts.

\vspace{2pt}
\noindent\textbf{Internal validity.}
Our measurements reflect each repository’s default branch at crawl time; configurations may differ before or after that snapshot. Some CI logic may be external (e.g., organization workflows, composite actions, or external CI systems), which can hide test execution and cause undercounting, so our labels reflect \emph{observable, in-repo} evidence. We scan workflow YAML plus referenced in-repo scripts/local actions when available and record SHAs for reproducibility.

In addition, a small fraction of repositories could not be cloned due to availability or timeouts, which may introduce selection bias. We de-duplicate repositories and allow multi-label assignment when multiple environment styles appear, avoiding forced single-style labels when evidence supports more than one.

We validate the CI instrumentation-test detection pipeline via a statistically grounded random sample (Section~\ref{sec:validation}). While the observed agreement provides an estimate of overall classification accuracy on the studied population, the sample size may not capture rare edge cases or subgroup-specific failure modes (e.g., uncommon CI providers, heavily indirect invocation through scripts/actions, or non-standard environment provisioning). Thus, residual misclassification risk may remain for atypical configurations.

Performance analysis is also limited to GitHub Actions runs available via the API at collection time; retention limits, missing runs, or API restrictions may reduce coverage and bias results. These signals are best-effort indicators and should not be over-interpreted for fine-grained differences.

Our evolution analysis (RQ2) mines default-branch history up to the cut-off date. Although we retrieve the necessary history for the repositories included in RQ2 (Section~\ref{sec:rq2-method}), history-based analyses can still be affected by factors such as rewritten history (e.g., force-pushes), incomplete linkage between commits and PR/issue discussions, and changes that occur outside the default branch.

\vspace{2pt}
\noindent\textbf{External validity.}
Our dataset is derived from public GitHub repositories selected using \texttt{stars:>50}, \texttt{fork:false}, and \texttt{archived:false} (Section~\ref{sec:repo-selection}). The \texttt{stars:>50} threshold biases the sample toward more visible and mature open-source projects; consequently, adoption rates and CI maturity signals observed in this study may be higher than those in the long tail of smaller or less-maintained repositories. Our sampling frame targets open-source projects and may not generalize to private/industry apps with different CI constraints, device access, compliance requirements, and proprietary tooling. Operational trade-offs derived from GitHub Actions telemetry may not transfer to other CI services or device-lab ecosystems with different failure modes, scheduling policies, and runtime profiles.

Moreover, repository-auditable artifacts may not fully expose environment setup when it is encapsulated in external marketplace actions or provided by self-hosted runners, which can limit the observability of execution-environment details even when instrumentation tests execute in CI.

\section{Conclusion}
\label{sec:conclusion}

We conducted a large-scale study of Android instrumentation testing in CI across 4{,}518 open-source repositories. Adoption is limited, often stopping at Gradle configuration without CI execution. Among CI adopters, \texttt{Community} and \texttt{Custom} dominate, while \texttt{GMD} and \texttt{Third Party} are rare. Execution environments change infrequently, but migrations typically move from \texttt{Custom} to \texttt{Community}, sometimes via brief mixed periods.  
Environment choice affects operational outcomes: \texttt{Community} offers the most reliable and efficient profile for push/PR gating, \texttt{Third Party} suits scheduled regression despite lower availability and higher cost, and \texttt{Custom} provides flexibility but experiences frequent reruns, reflecting workflow pressure. Our approach provides auditable, in-repo evidence of CI execution and environment style, though externalized configurations may remain unobserved.

\vspace{3pt}
\noindent\textbf{Future work.}
Future research should expand support for reusable/composite CI abstractions, improve longitudinal reconstruction beyond default-branch snapshots, and add developer-feedback KPIs capturing when actionable signals appear and why delays occur. We also plan to connect environment styles to downstream quality and cost signals, and build lightweight tooling to help projects safely migrate from \texttt{Custom} to \texttt{Community} or \texttt{GMD} using profile-driven recommendations and guardrails.

\section*{Artifacts and Replicability.}
The complete replication package, including all data, scripts, and step-by-step instructions for reproducing the results, is publicly available on Figshare: \url{https://doi.org/10.6084/m9.figshare.30934721}.

\section*{Acknowledgment}
We acknowledge the support of the Natural Sciences and Engineering Research Council of Canada (NSERC), [RGPIN-2021-03969].

\balance
\bibliographystyle{unsrt}
\bibliography{paper.bib}

\end{document}